\documentclass[8.5pt,twoside,twocolumn]{article}
\usepackage[super,sort&compress,comma]{natbib} 
\usepackage[version=4]{mhchem}
\usepackage{times,mathptmx}
\usepackage[dvipsnames]{xcolor}
\usepackage{sectsty}
\usepackage{titlesec}
\usepackage{balance} 
\usepackage{tabularx,subfigure}
\usepackage{amsmath,amsfonts,graphicx} 
\usepackage{lastpage}
\usepackage[format=plain,justification=raggedright,singlelinecheck=false,font=small,labelfont=bf,labelsep=space]{caption} 
\usepackage{fancyhdr}
\usepackage{graphicx}
\usepackage{etoolbox,hyperref}

\usepackage{parskip}
\begin{document}
\thispagestyle{plain}
\fancypagestyle{plain}{
\renewcommand{\headrulewidth}{1pt}}
\renewcommand{\thefootnote}{\fnsymbol{footnote}}
\renewcommand\footnoterule{\vspace*{1pt}%
\hrule width 3.4in height 0.4pt \vspace*{5pt}} 
\setcounter{secnumdepth}{5}

\captionsetup{justification=justified, singlelinecheck=off} 

\makeatletter 
\def\paragraph{\@startsection{paragraph}{4}{10pt}{-1.25ex plus -1ex minus -.1ex}{0ex plus 0ex}{\normalsize\textit}} 
\renewcommand\@biblabel[1]{#1}            
\renewcommand\@makefntext[1]%
{\noindent\makebox[0pt][r]{\@thefnmark\,}#1}
\makeatother 
\renewcommand{\figurename}{\small{Fig.}~}
\sectionfont{\large}
\subsectionfont{\normalsize} 

\fancyfoot{}
\fancyfoot[RO]{\footnotesize{\sffamily{~\textbar  \hspace{2pt}\thepage}}}
\fancyfoot[LE]{\footnotesize{\sffamily{\thepage~\textbar}}}
\fancyhead{}
\renewcommand{\headrulewidth}{1pt} 
\renewcommand{\footrulewidth}{1pt}
\setlength{\arrayrulewidth}{1pt}
\setlength{\columnsep}{6.5mm}
\setlength\bibsep{1pt}

\twocolumn[
  \begin{@twocolumnfalse}
\noindent\Large{\textbf{FIRE-GNN: Force-informed, Relaxed Equivariance Graph Neural Network for Rapid and Accurate Prediction of Surface Properties}}
\vspace{0.6cm}

\noindent\normalsize{\textbf{
Circe Hsu,\textit{$^{1,2}$} Claire Schlesinger,\textit{$^{2}$} Karan Mudaliar,\textit{$^{2}$} Jordan Leung,\textit{$^{2}$}, Robin Walters,\textit{$^{2}$} and Peter Schindler\textit{$^{3,*}$}
}}\vspace{0.5cm}


\normalsize{
\noindent
The work function and cleavage energy of a surface are critical properties that determine the viability of materials in electronic emission applications, semiconductor devices, and heterogeneous catalysis. While first principles calculations are accurate in predicting these properties, their computational expense combined with the vast search space of surfaces make a comprehensive screening approach with density functional theory (DFT) infeasible. Here, we introduce FIRE-GNN (Force-Informed, Relaxed Equivariance Graph Neural Network), which integrates surface-normal symmetry breaking and machine learning interatomic potential (MLIP)-derived force information, achieving a twofold reduction in mean absolute error (down to 0.065 eV) over the previous state-of-the-art for work function prediction. We additionally benchmark recent invariant and equivariant architectures, analyze the impact of symmetry breaking, and evaluate out-of-distribution generalization, demonstrating that FIRE-GNN consistently outperforms competing models for work function predictions. This model enables accurate and rapid predictions of the work function and cleavage energy across a vast chemical space and facilitates the discovery of materials with tuned surface properties.
}
\vspace{0.5cm}
 \end{@twocolumnfalse}
  ]

\footnotetext{\textit{
Northeastern University, Boston, MA 02115, USA\\
$^{1}$~Department of Mathematics\\
$^{2}$~Khoury College of Computer Sciences\\
$^{3}$~Department of Mechanical and Industrial Engineering\\
$^{*}$~Corresponding author: \href{mailto:p.schindler@northeastern.edu}{p.schindler@northeastern.edu}}}

\newcommand{\ourmodel}{FIRE-GNN}  

\section{Introduction}
The surface properties of crystalline materials are critical for a wide range of applications, including semiconductors~\cite{vandeurzenUsingBothFaces2024}, electron-emission devices~\cite{snappSidewallSiliconCarbide2012,barikDevelopmentAlloyfilmCoated2013a,watanabeSecondaryElectronEmission2011a,vossInherentEnhancementElectronic2014a}, batteries~\cite{dubeyTailoringSurfaceEnergy2022,heMXeneSurfaceEngineering2025}, and fuel cells~\cite{wangEffectExchangeSurface2024,artyushkovaStructuretopropertyRelationshipsFuel2012,maHighlyStableActive2024}. Work function, the energy required to extract an electron from the surface of a material, and cleavage or surface energy, the energy required to split a bulk crystal along a given Miller-index plane, are two important properties that govern the electronic properties and structural stability of a given crystal slab. The stability of a surface is crucial in preventing oxidation and corrosion~\cite{chenApplicationsDensityFunctional2025,liEffectSurfaceSegregation2022,xieEffectSurfaceStoichiometry2024}, which is particularly critical in sectors such as aerospace~\cite{zhuAdvancementsCorrosionProtection2024}, marine~\cite{lawalCorrosionControlIts2024}, and biomedical systems~\cite{prestatCorrosionTitaniumSimulated2021}, where failure is highly detrimental. Moreover, the surface energy determines the equilibrium shape of nanoparticles through the Wulff construction~\cite{wulffXXVZurFrage1901}. The work function governs the contact barrier at material interfaces, which is critical in semiconductor devices~\cite{brattainSurfacePropertiesSemiconductors1957,margaritondoInterfaceStatesSemiconductor1999,liuApproachingSchottkyMott2018}. Materials with low work functions are essential for applications such as electron-emission devices and thermionic energy converters~\cite{leeOptimalEmittercollectorGap2012a,campbellProgressHighPower2021}. In contrast, high work function materials are required in electronic components that demand large contact barriers~\cite{shaoWorkFunctionElectron2021}, such as to suppress leakage currents. 

Computational methods like density functional theory (DFT) can predict these key surface properties of existing and newly proposed materials with high fidelity from first principles. This enables a computationally-guided discovery framework that significantly reduces the time and cost associated with experimental synthesis and testing. However, the space of relevant surface orientations is large, typically requiring consideration of a Miller index of up to three~\cite{xiaoHighIndexFacetHighSurfaceEnergyNanocrystals2020}, resulting in tens to hundreds of unique surfaces per bulk crystal. The vast structural search space, coupled with the significant computational demands of \textit{ab initio} techniques, renders a fully exhaustive discovery pipeline computationally prohibitive. Recently, powerful model architectures have been introduced based on graph neural networks (GNNs) that can encode both crystal and molecule structure information and use the encoding to train a convolutional neural network to rapidly predict materials properties~\cite{xieCrystalGraphConvolutional2018a}. However, current ML methods often suffer from limited generalization, poor physical accuracy, and data inefficiency.

Equivariant neural networks (ENNs), which are neural networks that incorporate physical symmetries as hard mathematical constraints, help to address these limitations. ENNs are well-suited for crystal property prediction because they can exploit the inherent symmetry of crystal structures and enforce it in latent space, leading to improved prediction accuracy. They also exhibit greater data efficiency than non-equivariant models~\cite{batzner_e3-equivariant_2022}, particularly valuable in materials science, where data is scarce due to the high cost of DFT calculations and the time-intensive nature of experimental measurements. Finally, equivariant networks are well suited for predicting tensorial properties since their latent features are naturally encoded as geometric tensors, which transform appropriately relative to the coordinate system, giving an advantage relative to networks utilizing only scalar features.

In crystalline systems, ENNs have been employed to predict various tensor properties such as electric~\cite{yan_space_nodate} and mechanical~\cite{wen_equivariant_2024,pakornchote_straintensornet_2023}, vector properties such as force information~\cite{yang_lightweight_2024, yuan_equivariant_2024}, scalar fields such as electron density~\cite{jorgensen_equivariant_2022}, and scalar properties~\cite{kaba_equivariant_nodate}.

Equivariant networks have been widely applied to the prediction of bulk properties in crystalline materials, but their application to crystalline surfaces remains relatively limited, with the exception of molecule adsorption energies on surfaces, as highlighted by the OC20 dataset~\cite{tran_open_2023}. Bulk crystals exhibit full 3D rotational symmetry, allowing for the use of widely available E(3)-equivariant model libraries. Surfaces, however, break symmetry along their normal direction, making fully E(3) equivariant models potentially mismatched for modeling surface systems. Instead, surface slabs are more appropriately described by 2D rotational symmetry, represented by SO(2), within the plane defined by their two periodic directions.  One option is to apply $\mathrm{SO}(2)$ equivariant models, however this discards the 3D symmetry in the full dataset, which contains cuts along many different planes. Alternatively, to effectively adapt $ E(3)$-equivariant architectures to such systems, it is possible to relax equivariance constraints or introduce symmetry-breaking features.

In this work, we design FIRE-GNN (Force-Informed, Relaxed Equivariance Graph Neural Network), an E(3) equivariant GNN with symmetry-breaking normal to the surface and with machine learning interatomic potential (MLIP)-derived force information to accurately predict the cleavage energy and work functions of a material's surface, as a rapid surrogate model to DFT-based computations. We provide evidence that relaxed or symmetry-broken equivariant models are necessary to capture the specific symmetries of surfaces, unlike for bulk crystal systems. After thorough benchmarking of various invariant and equivariant architecture choices, we perform comprehensive out-of-distribution generalization tests for a variety of different holdout test sets, such as holding out elements or space groups, finding that equivariant models outperform non-equivariant models across all dataset splits. Finally, we introduce physically relevant geometric node features by leveraging MLIPs to generate atomic force labels, which, when coupled with symmetry-broken equivariant networks, greatly outperform the previous state-of-the-art models.

\section{Methods}

In this section, we explain the surface property dataset, how it is split for out-of-distribution generalizability analysis, the ML model baseline, the considered invariant and equivariant GNN architectures for benchmarking, and lastly, the design of our \ourmodel{} model architecture that is tuned for surface property predictions and its training procedure. 

\subsection{Surface Property Dataset}\label{database}
Existing databases that report the surface energy, cleavage energy, and the work function are limited in scale and diversity. Surface energies and work functions of about 1,500 elemental crystal surfaces have been reported with a maximum Miller index of three~\cite{tranAnisotropicWorkFunction2019}, about 3,500 surface energies and work functions of binary Mg intermetallics with Miller indices up to one~\cite{shiSurfaceEmphasizedMultitask2024b}, and cleavage energies of about 3,000 intermetallic surfaces with Miller indices up to two~\cite{palizhatiPredictingIntermetallicsSurface2019b}. The models in this work are trained on our previously established surface property dataset that consists of cleavage energies and work functions of 33,631 slabs, more than an order of magnitude larger than previous datasets.\cite{schindler_discovery_2024} A maximum Miller index of one and 3,716 bulk crystals up to ternary compounds with a near-zero band gap (obtained from the Materials Project~\cite{hortonAcceleratedDatadrivenMaterials2025}) were considered. The cleavage energies and work functions were computed using a high-throughput DFT framework utilizing the PBE exchange-correlation functional. The cleavage energy, $E_\mathrm{cleavage}$, is calculated with the total energies of the bulk and unrelaxed slab, $E_\mathrm{bulk}$ and $E_\mathrm{slab}$ as
\[ E_\mathrm{cleavage} = \frac{E_\mathrm{slab}-n_\mathrm{bulk\cdot E_\mathrm{bulk}}}{2\cdot A}, \]
where $n_\mathrm{bulk}$ is the number of bulk unit cells contained in the slab, and $A$ is the surface area of the slab. The cleavage energy coincides with the surface energy in the case of slabs where the top and bottom surfaces are symmetrically equivalent.

The work function is defined as
\[ \phi = E_\mathrm{vacuum} - E_\mathrm{F}, \]
where $E_\mathrm{vacuum}$ is the electrostatic vacuum energy level a few Angstroms away from the surface, and $E_\mathrm{F}$ is the Fermi level inside the slab. A dipole correction is applied during DFT calculations to ensure a flat electrostatic energy level away from the slab.

The average work function and cleavage energy ($\pm$ standard deviation) are $3.92\pm 0.86$ eV and $100.6\pm 45.8$ meV$/\AA{}^2$, respectively. While all cleavage energies are unique, some slabs may not have a unique top and bottom work function, due to slab symmetry. This resulted in 58,332 unique work function values, which is less than double the number of slabs. This dataset is publicly available on Zenodo~\cite{schindlerWorkFunctionCleavage2024}.

\subsection{Dataset Splitting for Out-of-Distribution Generalization Assessment}\label{splits}

For studying the out-of-distribution generalization performance of selected models, we implemented four different dataset split methods with the \textit{MatFold} Python package~\cite{d.witmanMatFoldSystematicInsights2025}. Each split is grouped into training, validation, and test sets at a ratio of 70-20-10. In the easiest split, \verb|Struc|, the dataset is partitioned according to the slabs' parent bulk material structure. This is to avoid data leakage between training and test sets because slabs generated from the same crystal structure, while structurally unique, can be structurally and chemically very similar. We also consider three additional splits with more challenging generalization conditions: \verb|Elem|, \verb|SPG|, and \verb|PTG|. For the elemental split, \verb|Elem|, five random elements are selected to form the holdout test set (Yb, Zn, Ho, W, Tc), while ensuring the size of the test set is 10\% of the total dataset. All slabs containing these elements are then placed in the test set, with the remaining slabs being split into training and validation sets, ensuring that the parent crystal structure of each slab is only present in either the training or the validation set. A similar process is used for splits based on the space group, \verb|SPG|, and periodic table group, \verb|PTG|. The unique test set labels are Pm$\bar{3}$m, P4/nbm, C222, P$\bar{6}$m2, P6/mmm, and periodic table groups 2, 17, respectively.

\subsection{Random Forest Model Baseline}\label{rfbaseline}
As a baseline, we utilize our previously developed random forest (RF) model, which achieves DFT-level accuracy (mean absolute error of $\sim 0.1$ eV) on the work function task~\cite{schindler_discovery_2024}. This model, \verb|wfrfmodel|, is publicly available on GitHub~\cite{schindlerWorkFunctionRandom2025} and can be easily installed with pip or uv. Compared to the deep learning models explored in this paper, \verb|wfrfmodel| utilizes a custom featurization procedure in which the top (bottom) three layers of the top (bottom) surface are extracted from the slab, and physically-relevant atomic features (e.g., electron affinity) are generated for the extracted atoms. This model has previously been combined with an \textit{ab initio} photoemission model~\cite{antoniukGeneralizableDensityFunctional2020b} to discover new ultra-bright photocathode materials~\cite{antoniukNovelUltrabrightAirStable2021a}.

\subsection{Benchmarking of Invariant and Equivariant Graph Neural Networks}

\subsubsection{Background}\label{inv-vs-equiv}
\textit{Geometric GNNs} are particularly well-suited for atomistic systems, where atoms are represented as nodes and edges are formed using distance-based cutoffs~\cite{velickovic_graph_2018,duval_hitchhikers_2024}. This representation enables message passing between neighboring atoms to capture local interactions. Node features typically encode atom types as learnable embeddings, while edge features represent geometric relationships such as displacement vectors or interatomic distances. A key challenge in incorporating coordinate information is the absence of a canonical reference frame. Geometric GNNs often address this by adopting a coordinate-free formulation, using pairwise distances or relative positions as edge features, thereby ensuring invariance to translations and rotations. However, invariant features are limited since they discard directional and orientation-dependent information. Equivariant neural networks provide a more suitable alternative, which avoids this limitation. 

Equivariance is the commutativity of functions and group actions. Considering a function $f\colon A \rightarrow A$ and a group action $G$ on the set $A$, then $f$ is equivariant if 
\[  g f(a) = f(g  a)\quad \forall a\in A, g\in G.\] 
Equivariance is a desirable property in neural network design, as it enables models to be robust to transformations such as rotations or reflections of the input. In materials systems, for instance, a 180-degree rotation of a surface slab does not alter intrinsic scalar properties like the cleavage energy. However, non-equivariant networks must explicitly learn this invariance from data. In contrast, equivariant neural networks inherently preserve the behavior of such non-orientation-dependent properties while correctly transforming orientation-dependent outputs. For example, when a slab is rotated by 180 degrees, an equivariant model will maintain the predicted cleavage energy but correctly swap the top and bottom work function values to reflect the change in orientation.

\subsubsection{Benchmark Models}

For baselines on our cleavage energy and work function dataset, we use Graph Attention Networks~\cite{velickovic_graph_2018} (GATs), which are GNNs that utilize an attention mechanism between nodes and integrate the attention value in the convolution between nodes. Another baseline is the well-established Crystal Graph Convolutional Neural Network~\cite{xie_crystal_2018} (CGCNNs), which is a GNN that utilizes the distance between atoms as additional features in the convolution. 

We also test the EquiformerV2 model, which is an E(3)-equivariant graph neural network based on a transformer architecture for atomic interaction modeling~\cite{liao_equiformer_2023}, which has achieved state-of-the-art in relevant tasks. It encodes atom types as irreducible representations of SO(3), combined with radial basis functions of interatomic distances, and employs efficient Spherical Channel Network (eSCN) convolutions to reduce tensor-product complexity from $O(L_{max}^6)$ to $O(L_{max}^3)$, enabling higher angular degrees (up to 8). EquiformerV2 employs equivariant graph attention mechanisms with attention re-normalization, separable $S^2$ activations, and degree-aware layer normalization, enhancing stability and expressivity. Final scalar or vector projections are aggregated to predict target properties.

Another recent model we benchmark is the iComFormer, which is an SE(3)-invariant transformer for crystalline materials that incorporates periodicity at the graph-construction stage~\cite{yan_complete_2024}. Each atom is linked to its nearest periodic images through the shortest lattice translation vectors. The resulting periodic graph encodes interatomic distances and bond–lattice angles within a cutoff radius, expanded through radial and angular basis functions, and self-attention acts exclusively on these scalar invariants. This design makes iComFormer effective for predicting scalar material properties. eComFormer builds on the same framework but augments the scalar representation with SO(3)-equivariant vector channels. These vector features carry displacement information that transforms consistently under rotation, and a mixed-attention mechanism combines scalar attention with tensor-product updates of vector channels. This coupling allows eComFormer to capture anisotropic effects while preserving full rotational equivariance.

Lastly, we evaluate the performance of Steerable Equivariant GNNs (SEGNNs), which are GNNs that utilize spherical harmonics as the input, output, and weights of the graph.  Spherical harmonics are steerable such that under rotations in $O(3)$ they transform equivalently via Wigner-D matrices. SEGNNs are well-adapted to our task since they allow for more complex non-linear messages to be passed between nodes. There are a few variants of the message passing operator in SEGNNs: message passing (SEGNN-MP), convolution (SEGNN-Conv), and transformer (SEGNN-Trans). These variations are similar to message passing~\cite{brandstetter_geometric_2022}, convolutional~\cite{brandstetter_geometric_2022}, and graph attention~\cite{fuchsSE3Transformers3DRotoTranslation2020} neural networks, respectively. We found that the SEGNN-MP variant achieved the best performance on our surface dataset (see Supplementary Section 2 and Supplementary Tables 1 and 2). Throughout the remainder of this manuscript, SEGNN refers specifically to SEGNN-MP.

\subsection{\ourmodel{} Model Architecture Design}
\begin{figure}[h]
    \centering\includegraphics[width=0.75\linewidth]{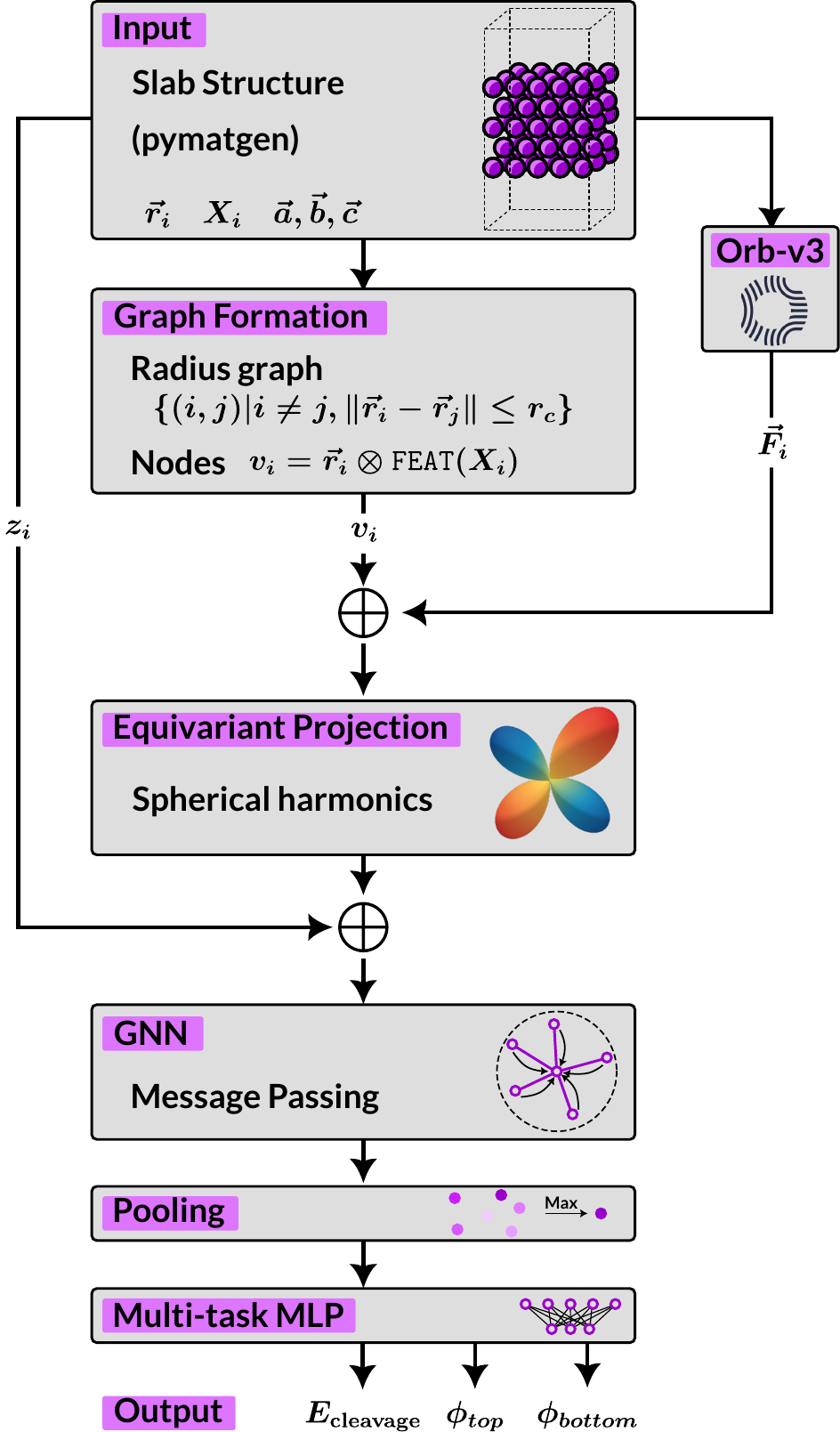}
    \caption{Overview of our final model architecture of \ourmodel. The input structure is represented as a pymatgen object that contains the atomic coordinates $\vec{r}_i$, atomic species $X_i$, and the lattice vectors $\vec{a}$, $\vec{b}$, and $\vec{c}$. The radius graph nodes are built through the concatenation of the atomic coordinates and the featurized atomic species (see Section \ref{elementalnodeablation}). The Orb-v3 MLIP features are then concatenated at the node level, followed by the equivariant projection. Symmetry breaking is achieved through the concatenation of the Cartesian $z_i$ coordinates before the GNN message passing layers. After pooling, a multi-task multi-layer perceptron (MLP) is used to predict the three target labels.}
    \label{fig:architecture}
\end{figure}

We designed a new model architecture by incorporating symmetry-breaking and physically relevant geometric force information into SEGNN, as explained in the next section. We will refer to this model as Force-Informed, Relaxed Equivariance GNN (\ourmodel{}) for the remainder of the manuscript. The model architecture is illustrated in detail in Fig.~\ref{fig:architecture}.

\subsubsection{Symmetry-broken Equivariance and Physically Relevant Geometric Features}\label{symbreak-orbv3}
As surfaces break the full E(3) symmetry along the direction normal to the surface, we incorporate the $z$-coordinate as a node-level feature, breaking the full SO(3) symmetry of the model and reducing the symmetry to SO(2). This better aligns the model symmetry with the task symmetry.

Foundational interatomic potentials, also referred to as universal MLIPs, have emerged as a powerful tool by being trained on extensive, high-fidelity materials datasets that cover a broad range of chemical compositions and structural motifs. These models enable accurate and transferable predictions of energies and atomic forces across diverse material systems, offering a “universal” applicability that transcends individual chemistries or material classes while being orders of magnitude faster than DFT. Here, we leverage the state-of-the-art Orb-v3 pretrained MLIP model\cite{rhodes_orb-v3_2025} to augment our surface database with physically relevant atomic force information, providing additional geometric signal to aid learning. The generated force information is concatenated on the node level of the graphs.

\subsection{Model Training}\label{modeltraining}
The models were trained on the training sets from the splits described in Section \ref{splits}. Convergence of models (number of epochs) was evaluated based on the validation set errors. Lastly, the final performance was evaluated on the test sets. Hyperparameters were set at previously established heuristics and are described in more detail in  Supplementary Section 1. To assess model performance, we utilize domain standard metrics, \textit{Mean Absolute Error} (MAE) and \textit{Mean Absolute Percentage Error} (MAPE). The MAE measures the average magnitude of the absolute errors between predicted $\hat{y}_i$ and true values $y_i$, and is defined as
\[
\text{MAE} = \frac{1}{n} \sum_{i=1}^{n} |y_i - \hat{y}_i|
\]
where  $n$ is the number of samples. The MAPE is defined as
\[
\text{MAPE} = \frac{100\%}{n} \sum_{i=1}^{n} \left| \frac{y_i - \hat{y}_i}{y_i} \right|.
\]

\section{Results and Discussions}

\subsection{Benchmarking of Invariant and Equivariant Models}
First, we compare several invariant and equivariant models as an initial baseline to determine which architecture is best-suited for surface property prediction. The MAEs for the mean baseline (a model which predicts the training dataset mean), the RF model\cite{schindler_discovery_2024}, GAT, CGCNN, Equiformer-v2, iComformer, eComformer, SEGNN, and \ourmodel{} for the three target tasks (cleavage energy, work function top, and work function bottom) are compiled in Table \ref{tab:MAE_results} (and the MAPEs are listed in Supplementary Table 3). Here, we discuss the performance of all benchmarking models without modifications (i.e., without symmetry-breaking or physically relevant force feature addition). GAT and the basic CGCNN models perform poorly on both cleavage energy and work function predictions (only a $\sim 30\%$ reduction in error compared to the mean baseline). The remaining standard invariant and equivariant models have uniformly good performance on the cleavage energy tasks, while the work function prediction errors are considerably larger due to the orientation-dependent nature of the work function (top vs. bottom). The best-performing model on the cleavage energy prediction is the iComformer model with a MAE of 3.2 meV$/\AA{}^2$. Notably, the basic Equiformer-v2 model exhibits a significantly larger error on the cleavage energy task compared to Comformer and SEGNN (MAE about three times larger), with significantly greater computational overhead. Moreover, Equiformer-v2 has a $\sim 50\%$ higher MAE on the work function task compared to iComformer. Multi-task learning is applied to SEGNN benchmarks (see column ``Multi'' in Table \ref{tab:MAE_results}), with a slight improvement in work function performance at the expense of doubling the MAE on the cleavage energy prediction task, which is expected due to balancing of the multi-loss optimization. For Comformer models, interestingly, the basic invariant variation significantly outperforms the equivariant variation on the work function task. We hypothesize that equivariance may not be beneficial for the orientation-dependent work function property and/or that model optimization in the equivariant version is more challenging. 

\begin{table*}[h]
\centering
\begin{tabular}{l|ccc|ccc|}
                 & \multicolumn{3}{c|}{Architecture} & \multicolumn{3}{c|}{MAE Train/Test}                                                                                          \\ \cline{2-7} 
Model           & Multi & Sym-break & MLIP & \multicolumn{1}{c|}{$E_\mathrm{cleavage} \; (\mathrm{meV}/ \AA{}^2)$} & \multicolumn{1}{c|}{$\phi_\mathrm{top}$ (eV)} & $\phi_\mathrm{bottom}$ (eV) \\ \hline
Mean Baseline & - & - & - & \multicolumn{1}{c|}{-- / 32.0}                                     & \multicolumn{1}{c|}{-- / 0.667}                     & -- / 0.632                        \\
RF & - & - & - & \multicolumn{1}{c|}{--}                                   & \multicolumn{1}{c|}{0.040 / 0.124}    & 0.040 / 0.125     \\
\hline
GAT & - & - & - & \multicolumn{1}{c|}{21.0 / 20.1} & \multicolumn{1}{c|}{0.502 / 0.522} &  \multicolumn{1}{c|}{0.515 / 0.526}\\
CGCNN & - & - & - & \multicolumn{1}{c|}{16.9 / 23.4} & \multicolumn{1}{c|}{0.480 / 0.492} &  \multicolumn{1}{c|}{0.482 / 0.483}\\
 & - & $\checkmark$ & - & \multicolumn{1}{c|}{18.9 / 16.1} & \multicolumn{1}{c|}{0.359 / 0.393} &  \multicolumn{1}{c|}{0.367 / 0.420}\\
\hline
Equiformer-v2 & - & - & - & \multicolumn{1}{c|}{6.5 / 9.3} & \multicolumn{1}{c|}{0.184 / 0.337} & \multicolumn{1}{c|}{0.213 / 0.383}\\
 & - & $\checkmark$ & - & \multicolumn{1}{c|}{3.6 / 5.1} & \multicolumn{1}{c|}{0.082 / 0.172} & \multicolumn{1}{c|}{0.062 / 0.163}\\
\hline
iComformer & - & - & - & \multicolumn{1}{c|}{0.6 / \textbf{3.2}} & \multicolumn{1}{c|}{0.041 / 0.204} & \multicolumn{1}{c|}{0.041 / 0.204}\\
 & - & $\checkmark$ & - & \multicolumn{1}{c|}{0.6 / \textbf{3.2}} & \multicolumn{1}{c|}{0.118 / 0.122} & \multicolumn{1}{c|}{0.118 / 0.115}\\
eComformer & - & - & - & \multicolumn{1}{c|}{0.6 / 3.3} & \multicolumn{1}{c|}{0.209 / 0.265} & \multicolumn{1}{c|}{0.209 / 0.270}\\
 & - & $\checkmark$ & - & \multicolumn{1}{c|}{0.7 / 3.4} & \multicolumn{1}{c|}{0.100 / 0.130} & \multicolumn{1}{c|}{0.100 / 0.134}\\
\hline
SEGNN & - & - & - & \multicolumn{1}{c|}{0.9 / 3.8} & \multicolumn{1}{c|}{0.083 / 0.287} &  \multicolumn{1}{c|}{0.085 / 0.283}\\
 & $\checkmark$ & - & - & \multicolumn{1}{c|}{6.9 / 7.9} & \multicolumn{1}{c|}{0.083 / 0.259} & \multicolumn{1}{c|}{0.082 / 0.254}\\
 & $\checkmark$ & $\checkmark$ & - & \multicolumn{1}{c|}{6.3 / 9.3} & \multicolumn{1}{c|}{0.067 / 0.252} & \multicolumn{1}{c|}{0.067 / 0.252}\\
\hline
\ourmodel{} (ours) & $\checkmark$ & $\checkmark$ & $\checkmark$ & \multicolumn{1}{c|}{4.0 / 6.1} & \multicolumn{1}{c|}{0.023 / \textbf{0.065}} & \multicolumn{1}{c|}{0.024 / \textbf{0.065}}\\
\end{tabular}
\caption{MAEs for various ML architectures for training/test sets of the \texttt{structureid} dataset split. Applied architecture choices are indicated by a tick mark. The mean baseline model predicts the average of the target label across the training and validation datasets for all slabs in the test set. RF model taken from our previous work~\cite{schindler_discovery_2024}, rerun on the split. The lowest error metrics are highlighted in bold for each of the three target labels.}
\label{tab:MAE_results}
\end{table*}

\subsection{Improved Performance through Symmetry-broken Equivariance}
\begin{figure}[h]
    \centering\includegraphics[width=1\linewidth]{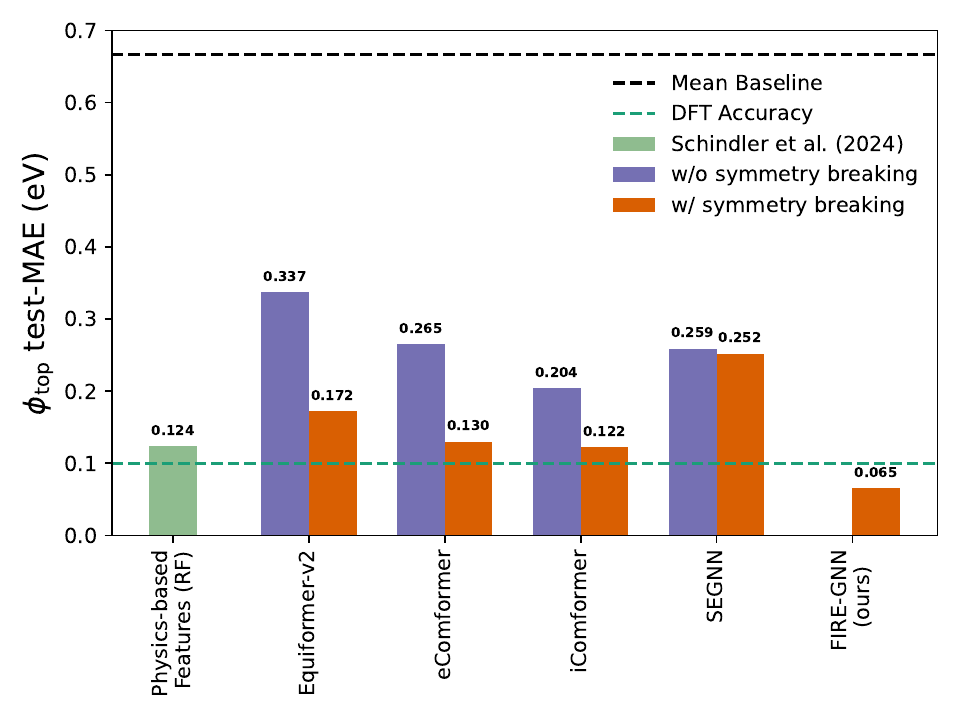}
    \caption{MAE performance metric for models with and without symmetry breaking. The mean baseline model predicts the average of the target label across the training and validation datasets for all slabs in the test set.}
    \label{fig:MAE_WF}
\end{figure}

As discussed in the Introduction, bulk crystals exhibit full three-dimensional rotational symmetry, while surfaces break symmetry along their normal direction and are instead more appropriately described by two-dimensional rotational symmetry, SO(2), within the periodic plane. Hence, we hypothesize that explicitly breaking symmetry in the $z$ direction will improve model performance for surface systems. To evaluate this, we tested the impact of $z$-direction symmetry breaking across several invariant and equivariant models. The results, summarized in Table \ref{tab:MAE_results} and visualized for the work function task in Fig.~\ref{fig:MAE_WF}, show consistent improvements, indicating that this modification significantly enhances performance and allows the networks to distinguish between top and bottom surfaces. Interestingly, such an improvement is not observed for the cleavage energy task (except for Equiformer-v2), as shown in Supplementary Figure 1. This indicates that the cleavage energy task, which is a non-directional scalar property, does not require symmetry breaking.

\subsection{Effect of Elemental Node Representation}\label{elementalnodeablation}
We further examined the effect of the elemental node featurization on the elemental out-of-distribution model performance. For this, we studied the SEGNN model (with multi-property prediction and symmetry breaking) and evaluated the MAEs on the elemental holdout split. Three elemental featurizations are considered: 1) Just the atomic number, 2) the period and group numbers, and 3) the full set of elemental features utilized in CGCNN featurization. The MAEs, reported in Table \ref{tab:elementablation}, show a substantial reduction when moving from featurization approach 1 to 2 (approximately 30\% for cleavage energy and 50\% for work function). Notably, no further improvement is observed when extending from approach 2 to 3, which indicates that the periodic table period and group numbers already capture enough information for the model to learn the underlying chemistry.


\begin{table}[h!]
\begin{tabular}{l|cc|}
                & \multicolumn{2}{c|}{MAE Train/Test}                                                                                          \\ \cline{2-3} 
Feature Type           & \multicolumn{1}{c|}{$E_\mathrm{cleavage} \; (\mathrm{meV}/ \AA{}^2)$} & \multicolumn{1}{c|}{$\phi$ (eV)} \\ \hline
Atomic number & \multicolumn{1}{c|}{6.5 / 12.7}                                    & \multicolumn{1}{c|}{0.061 / 0.455}      \\
Period and group \#       & \multicolumn{1}{c|}{6.9 / 9.0}                                   & \multicolumn{1}{c|}{0.065 / 0.229}      \\
Full Features & \multicolumn{1}{c|}{7.1 / 8.8} & \multicolumn{1}{c|}{0.064 / 0.246}\\
\end{tabular}
\caption{Train and test MAE of ablation experiments performed with the SEGNN model (multi-property, with symmetry breaking, without MLIP-features) and elemental generalization dataset split to determine whether extended node feature information aids in extrapolation performance.}
\label{tab:elementablation}
\end{table}

\subsection{Performance of FIRE-GNN}
As motivated earlier, we augmented our surface database with atomic force information from the Orb-v3 uMLIPs, concatenated at the node level alongside the introduced symmetry breaking in the $z$-direction. The resulting architecture outperforms all other benchmarks, achieving nearly a twofold reduction in MAE compared to the previous best model that relied on hand-crafted elemental features~\cite{schindler_discovery_2024}. \ourmodel{} predicts all three properties at once and also exhibits a low MAE for the cleavage energy prediction task (6.1 meV/$\AA{}^2$). However, this does not outperform the iComformer model with symmetry breaking due to the balancing of the multi-loss optimization of the multi-property prediction capability of \ourmodel{}.

\begin{figure*}[h]
    \centering\includegraphics[width=1\linewidth]{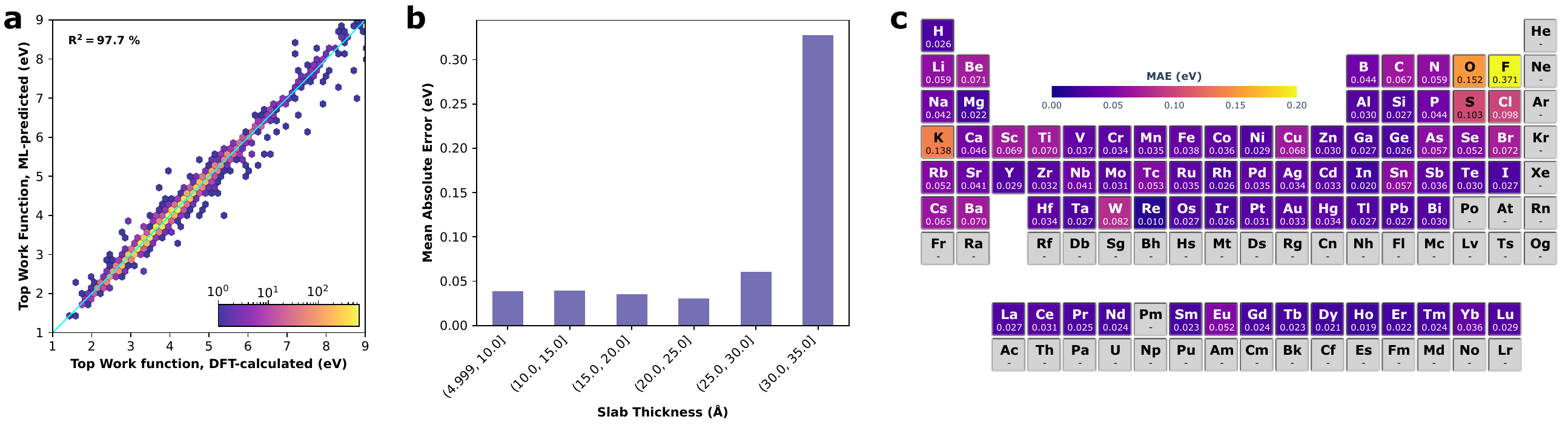}
    \caption{Performance assessment of \ourmodel{} predicting the top work function task. \textbf{a} Parity plot for the combined validation and test set. The color bar is displayed on a logarithmic scale. \textbf{b} MAEs for bins of different slab thicknesses. \textbf{c} Periodic table heatmap of MAEs averaged for all slabs where a specific element is present.}
    \label{fig:SEGNN-performance}
\end{figure*}

A detailed analysis of the performance of \ourmodel{} on the work function task is shown in Figure \ref{fig:SEGNN-performance}. The parity plot in Figure \ref{fig:SEGNN-performance}a demonstrates exceptional accuracy ($R^2=97.7\;\%$). Figure \ref{fig:SEGNN-performance}b plots the MAE binned by slab thickness, showing that the MAE is largely unaffected by the slab thickness up to a thickness of $30\;\AA{}$. For slabs thicker than $30;\text{\AA}$, we observe a marked increase in MAE (more than a factor of 6 larger), which we attribute to the growing graph size and the limited range of message passing, where information from the top of the slab can no longer propagate effectively to the bottom. Lastly, we show the MAEs averaged for all slabs where a specific element is present and visualize this as a heat map on the periodic table (Figure \ref{fig:SEGNN-performance}c). When fluorine is present in a slab the MAE is largest (0.371 eV), followed by oxygen (0.152 eV), potassium (0.138), sulfur (0.103), and chlorine (0.098). The larger prediction errors for fluorides and oxides may be attributed to their complex chemical bonding and behavior.

\subsection{Assessment of Out-of-Distribution Generealizability}
Lastly, we evaluated the out-of-distribution generalization performance of various invariant and equivariant models, as well as \ourmodel{} by holding out spacegroups, elements, or periodic table groups (cf.\ Section \ref{splits}). The test-MAEs for the four holdouts of the mean baseline, RF model, GAT, CGCNN, Equiformer-v2 (with and without symmetry breaking), iComformer and eComformer (with and without symmetry breaking), SEGNN, and \ourmodel{} are summarized in Table \ref{tab:MAE_OOD} (MAPEs are summarized in Supplementary Table 6). The generalization performance of SEGNN with message passing, convolution, and transformer layers was also studied and is summarized in Supplementary Tables 4 and 5. 

\begin{figure}[h]
    \centering\includegraphics[width=1\linewidth]{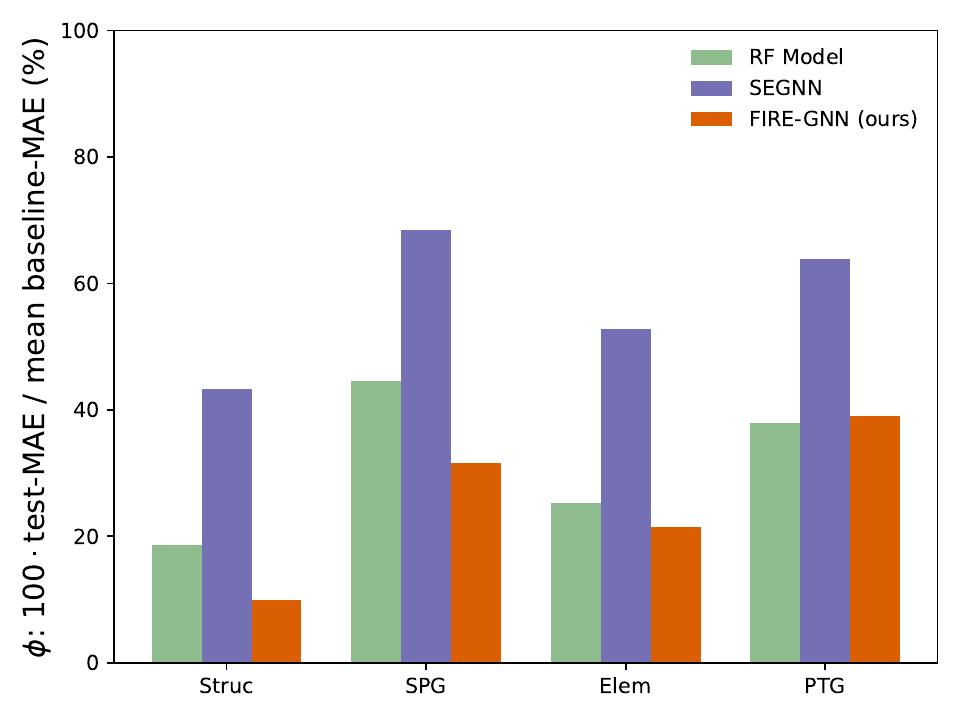}
    \caption{Test MAE relative to baseline MAE (in percent) for out-of-distribution generalization splits (\texttt{Struc}, \texttt{SPG}, \texttt{Elem}, and \texttt{PTG}) for RF model, SEGNN, and \ourmodel{} on the work function task.}
    \label{fig:MAE_WF_OOD}
\end{figure}

For the cleavage energy task, iComFormer with symmetry breaking achieves the best performance on the \verb|Struc|, \verb|SPG|, and \verb|Elem| holdouts, while SEGNN performs best on the \verb|PTG| holdout. For the work function task, \ourmodel{} attains the lowest errors across all holdouts except \verb|PTG|, where it performs on par with the RF model. This is illustrated in Figure \ref{fig:MAE_WF_OOD}, which shows test MAEs normalized to the mean baseline. \ourmodel{} consistently outperforms both SEGNN and RF on the \verb|Struc|, \verb|SPG|, and \verb|Elem| holdouts, and matches RF on \verb|PTG|. A corresponding plot for iComFormer with symmetry breaking on the cleavage energy task is provided in Supplementary Figure 2.

\begin{table*}[h!]
\centering
\begin{tabular}{l|ccc|cccccccc|}
        & \multicolumn{3}{c|}{Architecture} & \multicolumn{8}{c|}{MAE Test} \\ \cline{2-12} 
    & \multicolumn{3}{c|}{}     & \multicolumn{4}{c|}{$E_\mathrm{cleavage}\;(\mathrm{meV}/ \AA{}^2$)}                                                    & \multicolumn{4}{c|}{$\phi$ (eV)} \\ \cline{5-12} 
Model & Multi & Sym-break & MLIP     & \multicolumn{1}{c|}{Struc} & \multicolumn{1}{c|}{SPG} & \multicolumn{1}{c|}{Elem} & \multicolumn{1}{l|}{PTG}   & \multicolumn{1}{c|}{Struc.} & \multicolumn{1}{c|}{SPG} & \multicolumn{1}{c|}{Elem.} & \multicolumn{1}{l|}{PTG}  \\ \hline
Mean Baseline & - & - & -  & \multicolumn{1}{c|}{32.0}  & \multicolumn{1}{c|}{40.7}  & \multicolumn{1}{c|}{32.1} & \multicolumn{1}{c|}{41.7} & \multicolumn{1}{c|}{0.667}  & \multicolumn{1}{c|}{0.647}  & \multicolumn{1}{c|}{0.531} & \multicolumn{1}{c|}{0.796}                   \\
RF & - & - & -        & \multicolumn{1}{c|}{--}     & \multicolumn{1}{c|}{--}     & \multicolumn{1}{c|}{--}    & \multicolumn{1}{c|}{--}    & \multicolumn{1}{c|}{0.124}  & \multicolumn{1}{c|}{0.288}  & \multicolumn{1}{c|}{0.134} & \multicolumn{1}{c|}{\textbf{0.302}}                 
\\
 \hline
  GAT & - & - & -  & \multicolumn{1}{c|}{20.1} & \multicolumn{1}{c|}{21.9} & \multicolumn{1}{c|}{29.2} & \multicolumn{1}{c|}{38.4} & \multicolumn{1}{c|}{0.524} & \multicolumn{1}{c|}{0.512} & \multicolumn{1}{c|}{0.489} & \multicolumn{1}{c|}{0.686} \\
 CGCNN & - & - & - & \multicolumn{1}{c|}{23.4} & \multicolumn{1}{c|}{19.3} & \multicolumn{1}{c|}{39.0} &  \multicolumn{1}{c|}{62.4} & 
 \multicolumn{1}{c|}{0.488} & \multicolumn{1}{c|}{0.539} & \multicolumn{1}{c|}{0.484} & \multicolumn{1}{c|}{0.663} \\
 & - & $\checkmark$ & - & \multicolumn{1}{c|}{16.1} & \multicolumn{1}{c|}{25.0} & \multicolumn{1}{c|}{73.3} & \multicolumn{1}{c|}{109.5} &
\multicolumn{1}{c|}{0.407} & \multicolumn{1}{c|}{0.621} & \multicolumn{1}{c|}{0.447} & \multicolumn{1}{c|}{0.616}\\
\hline
 Equiformer-v2 & - & $\checkmark$ & -  & \multicolumn{1}{c|}{5.1} & \multicolumn{1}{c|}{12.6} & \multicolumn{1}{c|}{9.9} & \multicolumn{1}{c|}{23.5} & \multicolumn{1}{c|}{0.173} & \multicolumn{1}{c|}{0.390} & \multicolumn{1}{c|}{0.170} & \multicolumn{1}{c|}{0.612} \\
 \hline
  iComformer & - & - & -  & \multicolumn{1}{c|}{3.2} & \multicolumn{1}{c|}{8.2} & \multicolumn{1}{c|}{3.6} & \multicolumn{1}{c|}{ 
26.3} & 
 \multicolumn{1}{c|}{0.204} & \multicolumn{1}{c|}{0.427} & \multicolumn{1}{c|}{0.230} & \multicolumn{1}{c|}{0.481}\\
   & - & $\checkmark$ & -  & \multicolumn{1}{c|}{\textbf{3.1}} & \multicolumn{1}{c|}{\textbf{7.7}} & \multicolumn{1}{c|}{\textbf{3.5}} & \multicolumn{1}{c|}{26.2} & 
 \multicolumn{1}{c|}{0.148} & \multicolumn{1}{c|}{0.325} & \multicolumn{1}{c|}{0.136} & \multicolumn{1}{c|}{0.411}\\
  eComformer & - & - & -  & \multicolumn{1}{c|}{3.3} & \multicolumn{1}{c|}{8.7} & \multicolumn{1}{c|}{3.7} & \multicolumn{1}{c|}{26.3} & 
 \multicolumn{1}{c|}{0.268} & \multicolumn{1}{c|}{0.427} & \multicolumn{1}{c|}{0.230} & \multicolumn{1}{c|}{0.528}\\
   & - & $\checkmark$ & -  & \multicolumn{1}{c|}{3.4} & \multicolumn{1}{c|}{8.6} & \multicolumn{1}{c|}{3.9} & \multicolumn{1}{c|}{27.6} & 
 \multicolumn{1}{c|}{0.239} & \multicolumn{1}{c|}{0.339} & \multicolumn{1}{c|}{0.199} & \multicolumn{1}{c|}{0.416}\\
 \hline
  SEGNN & - & - & -  & \multicolumn{1}{c|}{3.8} & \multicolumn{1}{c|}{9.8} & \multicolumn{1}{c|}{6.4} & \multicolumn{1}{c|}{\textbf{18.7}} & \multicolumn{1}{c|}{0.285} & \multicolumn{1}{c|}{0.412} & \multicolumn{1}{c|}{0.275} & \multicolumn{1}{c|}{0.486} \\
\ourmodel{} (ours) & $\checkmark$ & $\checkmark$ & $\checkmark$  & \multicolumn{1}{c|}{6.1} & \multicolumn{1}{c|}{13.0} & \multicolumn{1}{c|}{8.3} & \multicolumn{1}{c|}{29.0} & 
 \multicolumn{1}{c|}{\textbf{0.066}} & \multicolumn{1}{c|}{\textbf{0.204}} & \multicolumn{1}{c|}{\textbf{0.114}} & \multicolumn{1}{c|}{0.310}\\
\end{tabular}
\caption{MAEs for various ML architectures for different out-of-distribution holdout test sets for both cleavage energy and work function tasks. Applied architecture choices are indicated by a tick mark. Holdouts: Bulk structure (\texttt{Struc}), space group  (\texttt{SPG}), elemental (\texttt{Elem}), and periodic table group (\texttt{PTG}). The lowest error metrics are highlighted in bold.}
\label{tab:MAE_OOD}
\end{table*}

\section{Conclusions}
In this work, we benchmarked a variety of invariant and equivariant neural network architectures and developed \ourmodel{}, a Force-Informed, Relaxed Equivariant Graph Neural Network architecture, which achieves state-of-the-art performance on work function prediction with a twofold improvement on prior state-of-the-art RF models. \ourmodel{} addresses major limitations of standard equivariant models by incorporating symmetry-breaking features along the slab $z$-normal axis to capture correct system equivariance. We further demonstrated that the addition of physically relevant atomic force information, derived from pretrained MLIPs, substantially improves model accuracy on property prediction tasks. \ourmodel{} displays exceptional out-of-distribution performance, and analysis of residual error demonstrates broad generalization across elements, with the sole exception of fluorine-containing slabs, likely due to its complex chemistry. 

Despite the strong performance of \ourmodel{}, there are a few limitations that provide opportunities for future work. In particular, the poor performance on structures with fluorine may be addressed by augmenting the dataset, reweighting the training samples, or changing the representations of the atom type.  Additionally, the performance of \ourmodel{} is worse for slabs thicker than 30 $\AA{}$ due to the limited receptive field of the GNN. This may be improved with deeper GNNs and rewiring methods used to combat oversmoothing issues in deeper GNNs.

The exceptional accuracy of \ourmodel{}, combined with orders-of-magnitude speedups over DFT, enables rapid exploration of materials space that was previously constrained by the high computational cost of DFT. While \ourmodel{} has the potential to accelerate screening via random search, we also foresee use of \ourmodel{} in \textit{de novo} design applications as a scoring function for generative modeling. Our model architecture may also be utilized in the future to predict materials properties of 2D materials, which naturally exhibit the same symmetry-breaking in the direction normal to the 2D material plane.


\section*{Data Availability Statement}

The dataset splits, final best model files, and the software to reproduce the predictions will be shared on Zenodo, \href{https://doi.org/10.5281/zenodo.16920229}{doi.org/10.5281/zenodo.16920229}, and GitHub, \href{https://github.com/d2r2group/FIRE-GNN}{github.com/d2r2group/FIRE-GNN}, after peer review.

\section*{Acknowledgments}

P.S. acknowledges the start-up funding from the Department of Mechanical and Industrial Engineering at Northeastern University. P.S. and R.W. acknowledge funding through the TIER1 and EAI Seed Grants. C.H. acknowledges funding from a Northeastern Undergraduate Research and Fellowships PEAK Experiences grant. C.S. acknowledges support through the NSF Graduate Research Fellowship. This work was completed in part using the \textit{Explorer} cluster, supported by Northeastern University’s Research Computing team. C.H. thanks Ariel Barr and Daniel Larson for helpful discussions.

\section*{Author Contributions}

C.\ H.\ contributed to Writing – Original Draft (lead), Writing – Review \& Editing (equal), Conceptualization (equal), Software (equal), and Investigation (equal). C.\ S.\ contributed to Writing – Original Draft (supporting), Writing – Review \& Editing (equal), Conceptualization (equal), Software (equal), and Investigation (equal). K.\ M.\ contributed to Conceptualization (equal), Software (equal), and Investigation (equal). J.\ L.\ contributed to Software (supporting) and Investigation (supporting). R.\ W.\ contributed to Writing – Review \& Editing (equal), Conceptualization (supporting), Resources (equal), Supervision (equal), and Funding acquisition (equal). P.\ S.\ contributed to Writing – Original Draft (supporting), Writing – Review \& Editing (equal), Conceptualization (supporting), Data Curation (lead), Visualization (lead), Resources (equal), Supervision (equal), and Funding acquisition (equal). All authors read and approved the final manuscript.

\clearpage

\footnotesize{
\bibliography{references.bib} 
\bibliographystyle{rsc} 
}

\end{document}



\sloppy

\maketitle


\section{Model Hyperparameters}
Our SEGNNs were constructed with 128 hidden features, 2 pre-pooling MLP layers, 2 post-pooling MLP layers, batchwise layer norm, and max pooling. For the SEGNN-MP and SEGNN-Conv models 3 graph layers were used, but for the SEGNN-Trans, only 1 was used for stability reasons. The learning rate was set to 5e-4 for the SEGNN-MP and SEGNN-Conv models and 5e-5 for the SEGNN-Trans model. We allowed up to order l=3 spherical harmonics in our models and used the Adam optimizer. Atoms were naively bonded to every other atom within 8 angstroms of itself.

For SEGNN with symmetry breaking and/or MLIP-features the hyperparameters were 384 hidden features, which is an increase compared to the baseline. We still used the same 3 graph layers, 2 pre-pooling MLP layers, and 2 post-pooling MLP layers. Batchwise layer normalization and max pooling were also kept the same from the baseline. However, the learning rate was increased to 8e-4, and the optimizer used was AdamW. Additionally, only up to order l=2 spherical harmonics were allowed in our models, and the radius for atomic bonds was set to 7 angstroms.

For our GAT and CGCNN models, we used 256 hidden features, 5 graph layers, and 3 MLP layers when predicting work function and 256 hidden features, 2 graph layers, and 2 MLP layers when predicting cleavage energy.

Both iComFormer and eComFormer utilized the following hyperparameters, which were selected to balance geometric resolution, model capacity, and training stability, consistent with the ComFormer design principles. A cutoff radius $r_c = 6$~\AA, an embedding dimension of 256, 8 attention heads, 8 transformer layers, a radial basis expansion size of 32, an angular basis size of 8, a batch size of 32, and a learning rate of $5\cdot 10^{-4}$ with cosine decay.

The EquiformerV2 31M model was fine-tuned on our dataset, fixing the hyperparameters at the pretrained values.

\section{Benchmarking SEGNN Layers}
We benchmarked the model performance of SEGNN with different layer choices. The train and test MAEs and MAPEs for the \emph{structureid} split are listed in \cref{tab:MAE_MP} and \cref{tab:MAPE_MP}. The out-of-distribution generalizability performance is shown in \cref{tab:MAE-OOD-MP} and \cref{tab:MAPE-OOD-MP} (MAE and MAPE, respectively).

\section{Cleavage Energy Prediction Analysis}
\begin{figure}[h]
    \centering\includegraphics[width=0.7\linewidth]{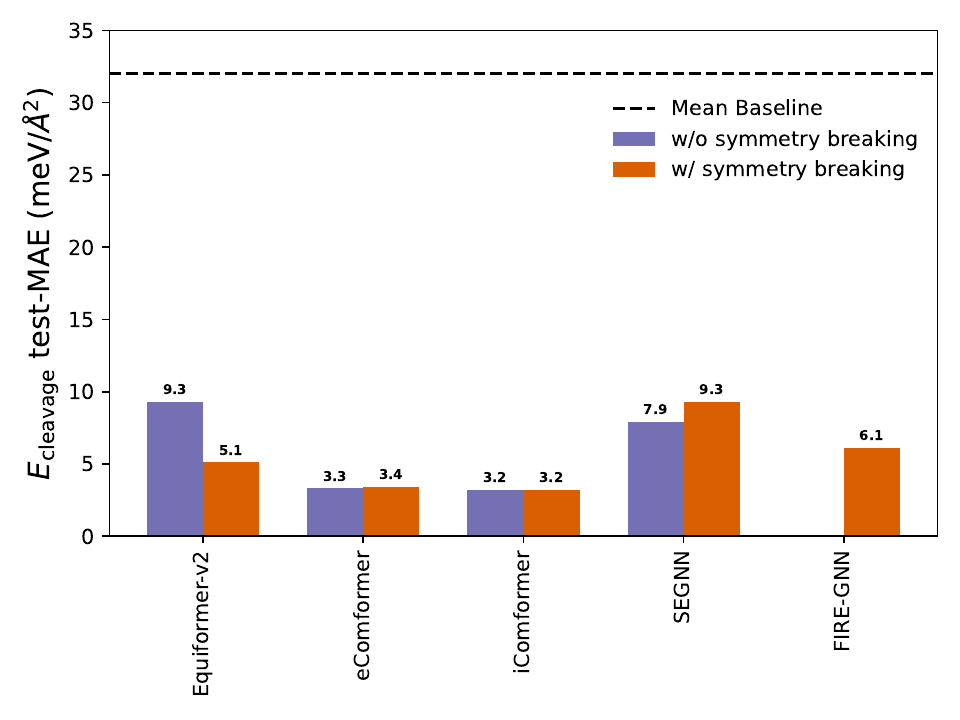}
    \caption{MAE performance metrics for models predicting the cleavage energy with and without symmetry breaking. }
    \label{fig:MAE_CE}
\end{figure}

Analogous to Fig.~2 in the main manuscript, the MAE performance metrics for various ML models with and without symmetry breaking is shown in \cref{fig:MAE_CE}.

\section{MAPE Metrics}
The main paper Tables focus on the MAE performance metric. Here we also show the MAPE of the models for the \emph{structureid} split and for the out-of-distribution generalizability performance in \cref{tab:MAPE_results} and \cref{tab:MAPE-OOD}, respectively.

\begin{table}[h!]
\centering
\begin{tabular}{l|ccc|}
                & \multicolumn{3}{c|}{MAE Train/Test}                                                                                          \\ \cline{2-4} 
Model           & \multicolumn{1}{c|}{$E_\mathrm{cleavage} \; (\mathrm{meV}/ \AA{}^2)$} & \multicolumn{1}{c|}{$\phi_\mathrm{top}$ (eV)} & $\phi_\mathrm{bottom}$ (eV) \\ \hline
SEGNN-MP & \multicolumn{1}{c|}{0.9 / \textbf{3.8}} & \multicolumn{1}{c|}{0.083 / \textbf{0.287}} &  \multicolumn{1}{c|}{0.085 / \textbf{0.283}}\\
SEGNN-Conv & \multicolumn{1}{c|}{1.4 / 5.5} & \multicolumn{1}{c|}{0.093 / 0.310} & \multicolumn{1}{c|}{0.096 / 0.301}\\
SEGNN-Trans & \multicolumn{1}{c|}{5.1 / 9.4} & \multicolumn{1}{c|}{0.264 / 0.381} & \multicolumn{1}{c|}{0.266 / 0.387}\\
\end{tabular}
\caption{MAE of SEGNN models with different layer choices: Message passing (MP), Convolution (Conv), and transformer (Trans). The lowest test errors are in bold.}
\label{tab:MAE_MP}
\end{table}

\begin{table}[h!]
\centering
\begin{tabular}{l|ccc|}
                & \multicolumn{3}{c|}{MAPE Train/Test (\%)}                                                           \\ \cline{2-4} 
Model           & \multicolumn{1}{c|}{$E_\mathrm{cleavage}$} & \multicolumn{1}{c|}{$\phi_\mathrm{top}$} & $\phi_\mathrm{bottom}$ \\ \hline
SEGNN-MP & \multicolumn{1}{c|}{1.15 / \textbf{4.23}} & \multicolumn{1}{c|}{2.11 / \textbf{7.39}} & \multicolumn{1}{c|}{2.24 / \textbf{7.14}} \\
SEGNN-Conv & \multicolumn{1}{c|}{1.85 / 6.27} & \multicolumn{1}{c|}{2.40 / 7.85} & \multicolumn{1}{c|}{2.51 / 7.69} \\
SEGNN-Trans & \multicolumn{1}{c|}{8.07 / 10.71} & \multicolumn{1}{c|}{7.01 / 9.76} & \multicolumn{1}{c|}{6.91 / 9.92}\\
\end{tabular}
\caption{MAPE of SEGNN models with different layer choices: Message passing (MP), Convolution (Conv), and transformer (Trans). The lowest test errors are in bold.}
\label{tab:MAPE_MP}
\end{table}

\begin{table}[h!]
\centering
\begin{tabular}{l|ccc|ccc|}
                 & \multicolumn{3}{c|}{Architecture} & \multicolumn{3}{c|}{MAPE Test (\%)}                                                                                          \\ \cline{2-7} 
Model           & Multi & Sym-break & MLIP & \multicolumn{1}{c|}{$E_\mathrm{cleavage}$} & \multicolumn{1}{c|}{$\phi_\mathrm{top}$} & $\phi_\mathrm{bottom}$ \\ \hline
Baseline & - & - & - & \multicolumn{1}{c|}{39.47}                      & \multicolumn{1}{c|}{17.23}               & 16.78                  \\
RF & - & - & - & \multicolumn{1}{c|}{--}                    & \multicolumn{1}{c|}{3.08}  & 3.2    \\
\hline
iComformer & - & - & - & \multicolumn{1}{c|}{3.45} & \multicolumn{1}{c|}{5.24} & \multicolumn{1}{c|}{5.22}\\
 & - & $\checkmark$ & - & \multicolumn{1}{c|}{3.55} & \multicolumn{1}{c|}{3.04} & \multicolumn{1}{c|}{2.96}\\
eComformer & - & - & - & \multicolumn{1}{c|}{3.64} & \multicolumn{1}{c|}{6.77} & \multicolumn{1}{c|}{6.96}\\
 & - & $\checkmark$ & - & \multicolumn{1}{c|}{3.65} & \multicolumn{1}{c|}{3.20} & \multicolumn{1}{c|}{3.45}\\
\hline
Equiformer-v2 & - & - & - & \multicolumn{1}{c|}{14.50} & \multicolumn{1}{c|}{7.51} & \multicolumn{1}{c|}{8.30}\\
 & - & $\checkmark$ & - & \multicolumn{1}{c|}{5.63} & \multicolumn{1}{c|}{4.14} & \multicolumn{1}{c|}{4.38}\\
\hline
CGCNN & - & - & - & \multicolumn{1}{c|}{25.75} & \multicolumn{1}{c|}{12.92} &  \multicolumn{1}{c|}{13.01}\\
CGCNN & - & $\checkmark$ & - & \multicolumn{1}{c|}{19.15} & \multicolumn{1}{c|}{10.61} &  \multicolumn{1}{c|}{11.67}\\
GAT & - & - & - & \multicolumn{1}{c|}{23.76} & \multicolumn{1}{c|}{13.71} & \multicolumn{1}{c|}{14.04} \\
\hline
SEGNN & - & - & - & \multicolumn{1}{c|}{4.23} & \multicolumn{1}{c|}{7.39} & \multicolumn{1}{c|}{7.14} \\
 & $\checkmark$ & - & - & \multicolumn{1}{c|}{9.1} & \multicolumn{1}{c|}{6.7} & \multicolumn{1}{c|}{6.5}\\
 & $\checkmark$ & $\checkmark$ & - & \multicolumn{1}{c|}{10.2} & \multicolumn{1}{c|}{6.3} & \multicolumn{1}{c|}{6.2}\\
\hline
SEGNN & $\checkmark$ & $\checkmark$ & $\checkmark$ & \multicolumn{1}{c|}{6.63} & \multicolumn{1}{c|}{1.57} & \multicolumn{1}{c|}{1.63}\\
\end{tabular}
\caption{MAPEs for various ML architectures for the test set of the \texttt{structureid} dataset split. Applied architecture choices are indicated by a tick mark. The mean baseline model predicts the average of the target label across the training and validation datasets for all slabs in the test set. RF model taken from our previous work, rerun on the split. The lowest error metrics are highlighted in bold for each of the three target labels.}
\label{tab:MAPE_results}
\end{table}

\begin{figure}[h]
    \centering\includegraphics[width=0.7\linewidth]{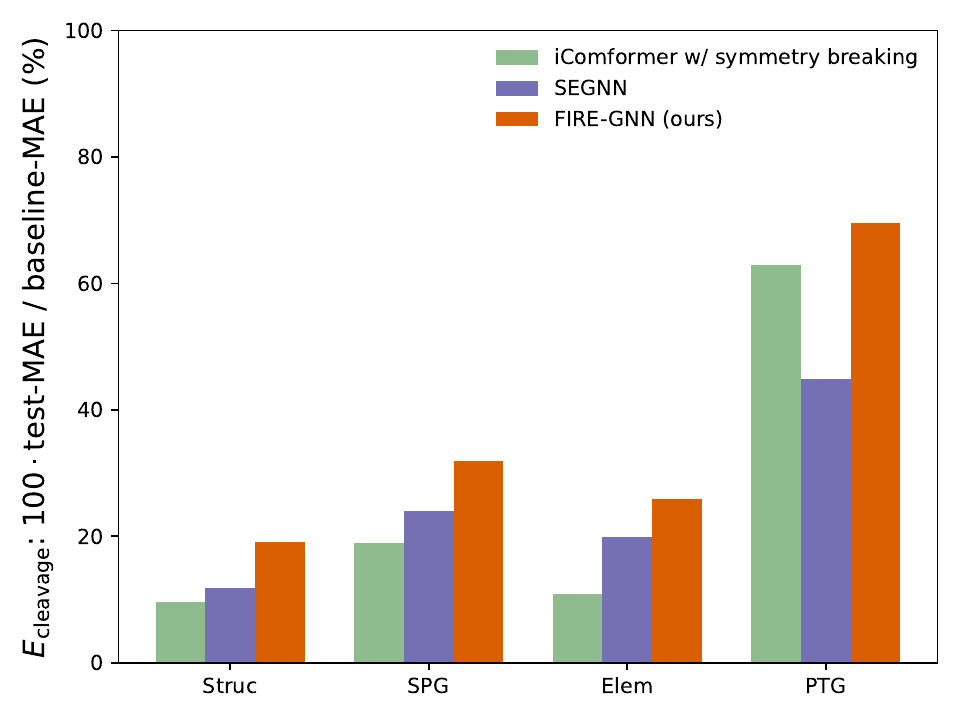}
    \caption{Test MAE relative to mean baseline MAE (in percent) for out-of-distribution generalization splits (Struc, SPG, Elem, and PTG) for iComformer with symmetry-breaking, SEGNN, and \ourmodel{} on the cleavage energy task. }
    \label{fig:CE_OOD}
\end{figure}

\begin{table}[h]
\centering
\begin{tabular}{l|cccccccc|}
         & \multicolumn{8}{c|}{MAE Test} \\ \cline{2-9} 
         & \multicolumn{4}{c|}{$E_\mathrm{cleavage}\;\mathrm{meV}/ \AA{}^2$}                                                    & \multicolumn{4}{c|}{$\phi$ (eV)} \\ \cline{2-9} 
Model    & \multicolumn{1}{c|}{Struc} & \multicolumn{1}{c|}{SPG} & \multicolumn{1}{c|}{Elem} & \multicolumn{1}{l|}{PTG}   & \multicolumn{1}{c|}{Struc.} & \multicolumn{1}{c|}{SPG} & \multicolumn{1}{c|}{Elem.} & \multicolumn{1}{l|}{PTG}  \\ \hline
 SEGNN-MP & \multicolumn{1}{c|}{3.8} & \multicolumn{1}{c|}{9.8} & \multicolumn{1}{c|}{6.4} & \multicolumn{1}{c|}{18.7} & \multicolumn{1}{c|}{0.285} & \multicolumn{1}{c|}{0.412} & \multicolumn{1}{c|}{0.275} & \multicolumn{1}{c|}{0.486} \\
 SEGNN-Conv & \multicolumn{1}{c|}{5.5} & \multicolumn{1}{c|}{10.9} & \multicolumn{1}{c|}{7.0} & \multicolumn{1}{c|}{18.0} &
 \multicolumn{1}{c|}{0.306} & \multicolumn{1}{c|}{0.443} & \multicolumn{1}{c|}{0.295} & \multicolumn{1}{c|}{0.507}\\
 SEGNN-Trans & \multicolumn{1}{c|}{9.4} & \multicolumn{1}{c|}{17.8} & \multicolumn{1}{c|}{9.8} & \multicolumn{1}{c|}{23.1} & 
 \multicolumn{1}{c|}{0.384} & \multicolumn{1}{c|}{0.509} & \multicolumn{1}{c|}{0.340} & \multicolumn{1}{c|}{0.577}\\

\end{tabular}
\caption{MAE of SEGNN models for different out-of-distribution splits with different layer choices: Message passing (MP), Convolution (Conv), and transformer (Trans). Holdouts: Bulk structure (struc), space group  (SPG), elemental (elem), and periodic table group (PTG).}
\label{tab:MAE-OOD-MP}
\end{table}

\begin{table}[h!]
\centering
\begin{tabular}{l|cccccccc|}
         & \multicolumn{8}{c|}{MAPE Test (\%)}\\ \cline{2-9} 
         & \multicolumn{4}{c|}{$E_\mathrm{cleavage}$}                                                                           & \multicolumn{4}{c|}{$\phi$} \\ \cline{2-9} 
Model    & \multicolumn{1}{c|}{Struc} & \multicolumn{1}{c|}{SPG} & \multicolumn{1}{c|}{Elem} & \multicolumn{1}{l|}{PTG}    & \multicolumn{1}{c|}{Struc.} & \multicolumn{1}{c|}{spgnum} & \multicolumn{1}{c|}{Elem.} & \multicolumn{1}{l|}{PTG} \\ \hline
SEGNN-MP & \multicolumn{1}{c|}{4.23} & \multicolumn{1}{c|}{11.06} & \multicolumn{1}{c|}{7.99} & \multicolumn{1}{c|}{67.50} & 
\multicolumn{1}{c|}{7.32} & \multicolumn{1}{c|}{10.70} & \multicolumn{1}{c|}{7.50} & \multicolumn{1}{c|}{12.24} \\
SEGNN-Conv & \multicolumn{1}{c|}{6.27} & \multicolumn{1}{c|}{13.65} & \multicolumn{1}{c|}{9.14} & \multicolumn{1}{c|}{64.44} & 
\multicolumn{1}{c|}{7.77} & \multicolumn{1}{c|}{11.72} & \multicolumn{1}{c|}{7.99} & \multicolumn{1}{c|}{12.58}
\\
SEGNN-Trans & \multicolumn{1}{c|}{10.71} & \multicolumn{1}{c|}{21.64} & \multicolumn{1}{c|}{11.72} & \multicolumn{1}{c|}{74.50} & 
\multicolumn{1}{c|}{9.84} & \multicolumn{1}{c|}{13.52} & \multicolumn{1}{c|}{9.08} & \multicolumn{1}{c|}{13.65}\\
\end{tabular}
\caption{MAPE of SEGNN models for different out-of-distribution splits with different layer choices: Message passing (MP), Convolution (Conv), and transformer (Trans). Holdouts: Bulk structure (struc), space group  (SPG), elemental (elem), and periodic table group (PTG).}
\label{tab:MAPE-OOD-MP}
\end{table}

\begin{table}[h!]
\centering
\begin{tabular}{l|ccc|cccccccc|}
        & \multicolumn{3}{c|}{Architecture} & \multicolumn{8}{c|}{MAPE Test (\%)} \\ \cline{2-12} 
    & \multicolumn{3}{c|}{}     & \multicolumn{4}{c|}{$E_\mathrm{cleavage}$}                                                   & \multicolumn{4}{c|}{$\phi$} \\ \cline{5-12} 
Model & Multi & Sym-break & MLIP     & \multicolumn{1}{c|}{Struc} & \multicolumn{1}{c|}{SPG} & \multicolumn{1}{c|}{Elem} & \multicolumn{1}{l|}{PTG}   & \multicolumn{1}{c|}{Struc.} & \multicolumn{1}{c|}{SPG} & \multicolumn{1}{c|}{Elem.} & \multicolumn{1}{l|}{PTG}  \\ \hline
Mean Baseline & - & - & -  & \multicolumn{1}{c|}{39.47}  & \multicolumn{1}{c|}{53.85}  & \multicolumn{1}{c|}{39.67} & \multicolumn{1}{c|}{128.34} & \multicolumn{1}{c|}{17.23}  & \multicolumn{1}{c|}{16.25}  & \multicolumn{1}{c|}{14.43} & \multicolumn{1}{c|}{20.28} \\
RF & - & - & -        & \multicolumn{1}{c|}{--}     & \multicolumn{1}{c|}{--}     & \multicolumn{1}{c|}{--}    & \multicolumn{1}{c|}{--}     & \multicolumn{1}{c|}{3.08}   & \multicolumn{1}{c|}{7.12}   & \multicolumn{1}{c|}{3.62}  & \multicolumn{1}{c|}{\textbf{7.23}}            
\\
\hline
iComformer & - & - & -  & \multicolumn{1}{c|}{\textbf{3.45}} & \multicolumn{1}{c|}{8.62} & \multicolumn{1}{c|}{\textbf{4.74}} & \multicolumn{1}{c|}{97.01} & 
\multicolumn{1}{c|}{5.23} & \multicolumn{1}{c|}{9.87} & \multicolumn{1}{c|}{4.86} & \multicolumn{1}{c|}{11.1}\\
 & - & $\checkmark$ & -  & \multicolumn{1}{c|}{3.55} & \multicolumn{1}{c|}{\textbf{8.35}} & \multicolumn{1}{c|}{\textbf{4.74}} & \multicolumn{1}{c|}{98.60} & 
\multicolumn{1}{c|}{3.00} & \multicolumn{1}{c|}{8.23} & \multicolumn{1}{c|}{3.53} & \multicolumn{1}{c|}{9.30}\\
eComformer & - & - & -  & \multicolumn{1}{c|}{3.64} & \multicolumn{1}{c|}{9.09} & \multicolumn{1}{c|}{6.12} & \multicolumn{1}{c|}{107.59} & 
\multicolumn{1}{c|}{6.86} & \multicolumn{1}{c|}{11.29} & \multicolumn{1}{c|}{6.12} & \multicolumn{1}{c|}{12.81}\\
 & - & $\checkmark$ & -  & \multicolumn{1}{c|}{3.65} & \multicolumn{1}{c|}{8.96} & \multicolumn{1}{c|}{4.92} & \multicolumn{1}{c|}{105.82} & 
\multicolumn{1}{c|}{3.32} & \multicolumn{1}{c|}{7.80} & \multicolumn{1}{c|}{3.27} & \multicolumn{1}{c|}{9.74}\\
\hline
Equiformer-v2  & - & $\checkmark$ & -  & \multicolumn{1}{c|}{5.63} & \multicolumn{1}{c|}{14.01} & \multicolumn{1}{c|}{17.68} & \multicolumn{1}{c|}{71.27} & 
\multicolumn{1}{c|}{4.14} & \multicolumn{1}{c|}{10.24} & \multicolumn{1}{c|}{4.70} & \multicolumn{1}{c|}{10.63} \\
\hline
CGCNN & - & - & - & \multicolumn{1}{c|}{25.75} & \multicolumn{1}{c|}{24.19} & \multicolumn{1}{c|}{47.78} & \multicolumn{1}{c|}{152.92} & 
 \multicolumn{1}{c|}{12.97} & \multicolumn{1}{c|}{14.07} & \multicolumn{1}{c|}{12.95} & \multicolumn{1}{c|}{15.51} \\
CGCNN & - & $\checkmark$ & - & \multicolumn{1}{c|}{19.15} & \multicolumn{1}{c|}{29.94} & \multicolumn{1}{c|}{92.44} & \multicolumn{1}{c|}{287.51} &
\multicolumn{1}{c|}{11.14} & \multicolumn{1}{c|}{15.55} & \multicolumn{1}{c|}{12.62} & \multicolumn{1}{c|}{14.31}\\
GAT  & - & - & -  & \multicolumn{1}{c|}{23.76} & \multicolumn{1}{c|}{28.98} & \multicolumn{1}{c|}{39.79} & \multicolumn{1}{c|}{109.68} & 
\multicolumn{1}{c|}{13.87} & \multicolumn{1}{c|}{13.29} & \multicolumn{1}{c|}{12.58} & \multicolumn{1}{c|}{15.74} \\
\hline
SEGNN  & - & - & -  & \multicolumn{1}{c|}{4.23} & \multicolumn{1}{c|}{11.06} & \multicolumn{1}{c|}{7.99} & \multicolumn{1}{c|}{\textbf{67.50}} & 
\multicolumn{1}{c|}{7.32} & \multicolumn{1}{c|}{10.70} & \multicolumn{1}{c|}{7.50} & \multicolumn{1}{c|}{12.24} \\
SEGNN & $\checkmark$ & $\checkmark$ & $\checkmark$  & \multicolumn{1}{c|}{7.07} & \multicolumn{1}{c|}{13.97} & \multicolumn{1}{c|}{9.44} & \multicolumn{1}{c|}{85.40} & 
 \multicolumn{1}{c|}{\textbf{1.59}} & \multicolumn{1}{c|}{\textbf{5.15}} & \multicolumn{1}{c|}{\textbf{3.11}} & \multicolumn{1}{c|}{7.30}\\
\end{tabular}
\caption{MAPEs for various ML architectures for different out-of-distribution holdout test sets for both cleavage energy and work function tasks. Applied architecture choices are indicated by a tick mark. Holdouts: Bulk structure (Struc.), space group  (SPG), elemental (Elem.), and periodic table group (PTG). The lowest error metrics are highlighted in bold.}
\label{tab:MAPE-OOD}
\end{table}

\printbibliography